\documentclass[useAMS,usenatbib,onecolumn]{mn2e}

\usepackage{amsmath}
\usepackage{graphicx}

\newcommand\calE{{\cal E}}
\newcommand\bfcalE{\boldsymbol{\cal E}}

\title[Flux transport and interface dynamos]
  {The effects of flux transport on interface dynamos}
\author[J. Mason, D.W. Hughes and S.M. Tobias]
  {J.~Mason,$^{1,2}$\thanks{E-mail: jmason@flash.uchicago.edu; d.w.hughes@leeds.ac.uk; smt@maths.leeds.ac.uk.}   
D.W.~Hughes$^1$ and S.M.~Tobias$^1$ \\
  $^1$Department of Applied Mathematics, University of Leeds, Leeds LS2 9JT, UK \\
  $^2$Department of Astronomy and Astrophysics, University of Chicago, Chicago, IL 60637, USA}

\begin{document}
\date{\today}

\pagerange{\pageref{firstpage}--\pageref{lastpage}} \pubyear{2008}

\maketitle
\noindent {The definitive version is available at www.blackwell-synergy.com}
\label{firstpage}

\begin{abstract}
The operation of an interface dynamo (as has been suggested for the Sun and other stars with convective envelopes) 
relies crucially upon the effective transport of magnetic flux between two spatially disjoint generation regions. In the simplest 
models communication between the two regions is achieved solely by diffusion. Here we incorporate a highly simplified 
anisotropic transport mechanism in order to model the net effect of flux conveyance by magnetic pumping and by magnetic 
buoyancy. We investigate the influence of this mechanism on the efficiency of kinematic dynamo action. It is found that the 
effect of flux transport on the efficiency of the dynamo is dependent upon the spatial profile of the transport. Typically, 
transport hinders the onset of dynamo action and increases the frequency of the dynamo waves. However, in certain cases 
there exists a preferred magnitude of transport for which dynamo action is most efficient. Furthermore, we demonstrate the 
importance of the imposition of boundary conditions in drawing conclusions on the role of transport.
\end{abstract}


\section{Introduction}
\label{intro}

The solar magnetic field is maintained by the action of a hydromagnetic dynamo. The well-known eleven year solar activity 
cycle corresponds to a twenty-two year magnetic cycle, the origin of which lies deep within the solar interior (see, for example, 
Charbonneau 2005). The key to understanding the generation of the solar magnetic field lies in establishing the form of both the 
large- and small-scale fluid flows in the interior. Helioseismology, the study of acoustic fluctuations within the Sun, has 
provided a map of the solar differential rotation (Schou et al.~1998), which shows that the surface latitudinal dependence is 
largely maintained throughout the convection zone while the solar interior rotates as a solid body. These two profiles are 
matched in a thin, stably stratified layer of strong differential rotation (shear) at the base of the convection zone -- a region 
known as the tachocline. A discussion of the dynamics of the tachocline can be found in Tobias (2005) and Hughes, Rosner 
\& Weiss (2007). 

This region of pronounced shear is believed to be of paramount importance in the generation (via the so-called 
$\omega$-effect), storage and eventual instability of a strong toroidal magnetic field. The regeneration of poloidal field from 
toroidal field is less well understood. It is typically ascribed to the $\alpha$-effect of mean field electrodynamics, but the 
precise origin and location of this mechanism remain unclear. A number of possibilities have been suggested (see the review of 
Ossendrijver 2003). The most natural is to appeal to convective turbulence in the presence of rotation (Parker 1955; Steenbeck, 
Krause \& R\"adler 1966). There are however some difficulties for this approach both in the linear (Cattaneo \& Hughes 2006, 
Hughes \& Cattaneo 2008) and nonlinear regimes (see the reviews by Diamond, Hughes \& Kim 2005; Brandenburg \& 
Subramanian 2005). An alternative is to appeal to magnetic buoyancy instabilities of the tachocline field interacting with rotation 
(Schmitt, Sch\"ussler \& Ferriz-Mas 1996; Thelen 2000; see also Cline, Brummell \& Cattaneo 2003 for a related model). A 
third possibility places the generation of poloidal field well away from the tachocline, the assumed location for the generation of 
toroidal field. In this paradigm (Dikpati \& Charbonneau 1999; Dikpati \& Gilman 2001) poloidal flux is regenerated via the 
decay of
active regions at the solar surface (Babcock 1961; Leighton 1969). In these models meridional circulation is invoked in order to 
transport the new poloidal flux to the tachocline. This model too is not without difficulties (see the review by Tobias \& Weiss 
2007).

In this paper we shall consider models of dynamos operating at an interface between regions of turbulent rotating convection 
and velocity shear. The idea of an interface dynamo --- with neighbouring regions of $\alpha$- and $\omega$-effects --- was 
introduced by Parker (1993) expressly to avoid the difficulties caused by the strong suppression of turbulent magnetic diffusion 
by a weak mean field. In Parker's model, strong toroidal field is generated in the tachocline by radial shear. There, and only 
there, does it suppress the diffusion, leading naturally to confinement of the toroidal field within that layer. The $\alpha$-effect 
and a healthy turbulent diffusion operate relatively unimpeded  in the bulk of the convection zone. Parker suggested that 
provided the diffusivity is not suppressed too strongly, observed solar fluxes and the 22-year solar period could be 
accommodated. Parker's original kinematic model has been extended into the nonlinear regime to investigate explicitly the 
quenching not only of the turbulent diffusion but also of the $\alpha$-effect (Tobias 1996; Charbonneau \& MacGregor 1997; 
Markiel \& Thomas 1999).

A crucial ingredient for efficient interface dynamo action is the transport of magnetic flux between the two spatially disjoint 
generation regions. For this reason, dynamo action is, typically, more difficult to excite in such models than in those with 
spatially-coincident generation mechanisms (Parker 1993). Indeed, if such a transport mechanism is absent, this places a severe 
constraint on the applicability of interface dynamos (Dikpati et al.\ 2005). Within the Sun there are however a number of 
possible mechanisms that could transport magnetic field. For example, convective pumping (e.g.\ Drobyshevski \& Yuferev 
1974) is believed to transport flux downwards from the convection zone into the tachocline (e.g.\ Tobias et al.\ 1998, 2001; 
Dorch \& Nordlund 2001). Turbulent diffusion, which is believed to act throughout the convection zone but could be 
significantly reduced in the tachocline (Parker 1993), acts to reduce gradients in field. Mean flows, such as the meridional 
circulation (Choudhuri et al. 1995), which is observed to be polewards at and just below the solar surface but is undetermined 
at the base of the convection zone, will act to advect mean flux. The only transport mechanism that is unambiguously directed 
outwards is magnetic buoyancy, which is believed to play a key role in the emergence of active regions at the surface (see, for 
example, the reviews by Sch\"ussler 2005 and Hughes 2007). These competing effects may occur on different timescales and 
depend on the local strength of magnetic field. Indeed, the general picture we have of the solar dynamo is of the field being 
confined and maintained at the base of the convection zone, with occasional eruptions.

In this paper we consider the effect of incorporating non-diffusive flux transport in an interface dynamo model. In keeping with 
the simple yet illustrative nature of Parker's original model, we include magnetic transport via an effective velocity. Indeed, such 
a velocity arises naturally in mean field electrodynamics as the `$\gamma$-effect', resulting from the anti-symmetric part of the 
$\alpha$-tensor (R\"adler 1968). It should be noted that mean field electrodynamics is not without problems, particularly in the 
high magnetic Reynolds number regime applicable to the Sun. Nonetheless, our hope is that it is possible to capture some 
features of the interaction of generation, transport and diffusion of magnetic fields via a simple mean-field model. 

In this paper we shall investigate the role of the $\gamma$-effect on the onset of the dynamo instability for two 
kinematic $\alpha\omega$-dynamo models. A natural starting point is to extend Parker's 1993 model, in which the interface 
separates two spatially unbounded regions. It is though important to consider the more realistic case of vertically bounded 
domains, which is the aim of our second model. In \S2 we discuss those characteristics that are common to both models. The 
effect of introducing flux transport is investigated, for the two models, in \S3 and \S4. The possible implications for an 
interface dynamo in the Sun are discussed in \S5.

\section{Formulation of the Parker Interface dynamo}
\label{sec:common}

\begin{figure}
\center\resizebox{0.6\textwidth}{!}{ \includegraphics{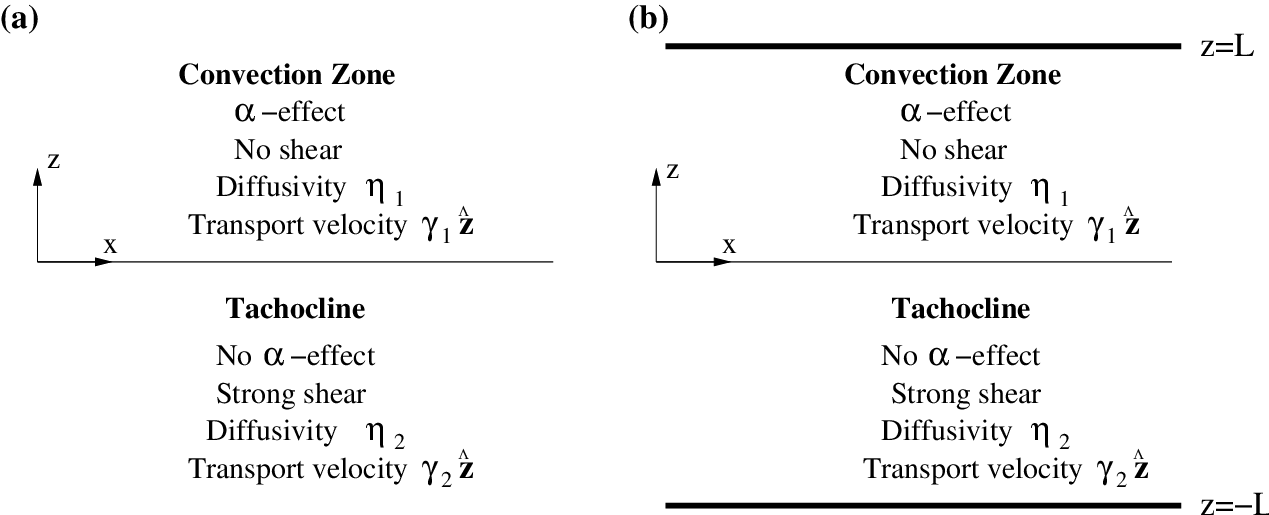}}
\caption{Geometry of the two models: (a) Model~I, (b) Model~II. We consider a local Cartesian geometry in the northern 
hemisphere, with the $z$-axis pointing radially outwards (upwards), the $x$-axis pointing towards the equator and the $y$-axis 
in the azimuthal direction. The models differ only by the $z$-extent of the domain.}
\label{fig:geometry}
\end{figure}

We consider the role of flux transport for two models of kinematic, $\alpha\omega$-interface dynamos in a 
local Cartesian geometry (see Figure~\ref{fig:geometry}). The region $z>0$ represents the convection zone (although there is 
no convection as such in our framework), wherein the $\alpha$-effect generates the poloidal field and there is no 
$\omega$-effect. The region $z<0$ models the tachocline, where the radial shear regenerates the toroidal field (the 
$\omega$-effect) and 
$\alpha=0$. The assumption of axisymmetry translates to solutions that are independent of the azimuthal direction, 
$y$ say, and we shall consider fields that are periodic in $x$.

The induction equation for the mean field $\boldsymbol{B}$ takes the form (see, for example, Moffatt 1978)
\begin{equation}
\label{eq:mf3} 
\frac{{\partial}{\boldsymbol{B}}}{{\partial}{t}}= \nabla
\times (\boldsymbol{U} \times \boldsymbol{B}) + \nabla
\times \bfcalE + \eta_0 \nabla^2 \boldsymbol{B},
\end{equation}
where $\boldsymbol{U}$ represents the mean flow, $\eta_0$ is the molecular magnetic diffusivity, and $\bfcalE$ is the mean 
electromotive force, which takes the form
\begin{equation}
\label{eq:mf4} 
\calE_i = \alpha_{ij}B_j +\beta_{ijk} \frac{\partial B_j}{\partial x_k} + \ldots.
\end{equation}
For simplicity we shall consider the effects of anisotropy only in the $\alpha_{ij}$ tensor, which we assume to take the form 
$\alpha_{ij} = \alpha \delta_{ij} - \epsilon_{ijk} \gamma_k$. We assume that $\beta_{ijk}=\beta \epsilon_{ijk}$; $\beta$ is 
then a turbulent magnetic diffusivity. Equation (\ref{eq:mf3}) now becomes
\begin{equation}
\label{eq:mf2}
\frac{{\partial}\boldsymbol{B}}{{\partial}{t}} = 
\nabla \times (\boldsymbol{U} \times \boldsymbol{B}) 
+ \nabla \times (\boldsymbol{\gamma} \times \boldsymbol{B}) +
\nabla \times (\alpha \boldsymbol{B}) -
\nabla \times \left( \eta \left( \nabla \times \boldsymbol{B} \right) \right),
\end{equation}
in which $\eta = \beta + \eta_0$.
We shall consider models in which $\eta$ may take different constant values in $z>0$ and $z<0$; thus, for the purpose 
of deriving the jump relations below, we do not take $\eta$ outside the derivative in the above equation.

The large-scale velocity $\boldsymbol{U}$ is chosen to represent the radial 
shear of the tachocline.  The effective velocity $\boldsymbol{\gamma}$ describes the radial
non-diffusive transport of magnetic flux between the two layers. We therefore take
\begin{equation}
\label{eq:pumping}
\boldsymbol{U}=U(z) \boldsymbol{\hat y}, \qquad  \boldsymbol{\gamma}=\gamma(z) \boldsymbol{\hat z},
\end{equation}
where $dU/dz=G H(-z)$, $\gamma(z)=\gamma_2+(\gamma_1-\gamma_2)H(z)$ and $H(z)$ is the Heaviside function. 
Here $G$, $\gamma_1$ and $\gamma_2$ are constants. We may think of the situation in which $\gamma_{i}<0$ ($i=1,2$) 
as the case in which magnetic pumping is the prevailing transport mechanism leading to a net transport radially inwards,
while the case in which the dominant transport effect is by magnetic buoyancy (and hence directed radially outwards) 
is modelled by $\gamma_i>0$. 

As the system is axisymmetric ($y$-independent), it is useful to decompose the field into poloidal and toroidal parts, 
$\boldsymbol{B}(x,z,t)=\nabla \times A \boldsymbol{\hat y} + B \boldsymbol{\hat y}$. We proceed with the convention that 
subscripts $1$ and $2$ refer to quantities in $z>0$ and 
$z<0$, respectively. Assuming that the shear is much stronger than the $\alpha$-effect in generating the toroidal field 
(the $\alpha\omega$-approximation), it follows from equation (\ref{eq:mf2}) that in $z>0$ the governing equations are 
\begin{equation}
\label{eq:dynamo_z>0}
\frac{{\partial}{A_1}}{{\partial}{t}} = {\alpha B_1
+\eta_1 \nabla^{2} A_1 - \gamma_1 \frac{{\partial}{A_1}}{{\partial}{z}}}, \qquad
\frac{{\partial}{B_1}}{{\partial}{t}} =
{\eta_1 \nabla^{2}B_1-\gamma_1 \frac{{\partial}{B_1}}{{\partial}{z}}},
\end{equation}
while in $z<0$ we have
\begin{equation}
\label{eq:dynamo_z<0}
\frac{{\partial}{A_2}}{{\partial}{t}} = {\eta_2 \nabla^{2} A_2 - \gamma_2 \frac{{\partial}{A_2}}{{\partial}{z}}}, \qquad
\frac{{\partial}{B_2}}{{\partial}{t}} =
{\eta_2\nabla^{2}B_2+G \frac{{\partial}{A_2}}{{\partial}{x}}-\gamma_2 \frac{{\partial}{B_2}}{{\partial}{z}}}.
\end{equation}
In both regions we seek wave-like solutions of the form 
\begin{equation}
\label{eq:solns}
A=\Re[a(z)e^{pt+\textrm{i}kx}], \qquad B=\Re[b(z)e^{pt+\textrm{i}kx}] ,
\end{equation}
where $p=\sigma+\textrm{i} \omega$ ($\sigma$, $\omega \in \Re$), $k$ is the wavenumber ($k>0$), and $a(z)$ and $b(z)$ are 
the 
(complex) amplitudes of the dynamo waves. Temporally growing solutions ($\sigma>0$), with period $T=2 \pi/\omega$,
propagate equatorwards when $\omega<0$ and towards the north pole when $\omega>0$. Both $A$ and $B$ are continuous. 
The solutions are matched at the interface $z=0$ by the 
jump conditions (derived by integrating equation (\ref{eq:mf2}) across the interface)
\begin{equation}
\label{eq:jump}
\Big[\frac{{\partial}{A}}{{\partial}{z}}\Big]_0=0,\qquad \Big[\eta \frac{{\partial}{B}}{{\partial}{z}}-
\gamma B \Big]_{0}=0,
\end{equation} 
where $[\cdot\cdot]_0$ denotes the jump in the specified quantity across $z=0$. 
Applying the boundary conditions in $z$, which will be described later, leads to a dispersion relation connecting the growth rate 
and wavenumber to the non-dimensional dynamo number ($D$, say), which measures the relative strength of induction to 
diffusion. As the definition of $D$ differs between the two models, we postpone its definition to the later sections.

To set the scene we now describe the spatial form of the transport mechanisms that we implement. As already described, the 
efficiency of an interface dynamo is crucially reliant on flux transport between the two generation layers. In Parker's model 
transport is solely by diffusion, with, in the simplest case, $\eta_1=\eta_2$. The intention of Parker's model was to show how a 
strong field could be confined to a thin layer beneath the convection zone owing to that field suppressing the effective 
diffusion, leading to $\eta_2 \ll \eta_1$.  We shall therefore investigate the role of the $\gamma$-effect in models with uniform 
diffusion ($\eta_1=\eta_2$) and in those with $\eta_2 / \eta_1 \ll 1$. We will consider the following illustrative cases:

\begin{itemize}
\item {\bf Bidirectional transport with uniform diffusion}\\
We consider first the case in which $\gamma_1=-\gamma_2=\gamma$ with $\gamma<0$, modelling pumping in $z>0$
and magnetic buoyancy in $z<0$. For simplicity we take $\eta_1=\eta_2$. Although one might expect stronger 
transport to increase the efficiency of the dynamo, we will show that this is not necessarily the case. We will then compare 
these results with the situation $\gamma>0$, which represents flux transport away from the interface in both generation layers 
and hence may naively be expected to yield the least efficient dynamo. 

\item {\bf Unidirectional transport with uniform diffusion}\\
Next we consider $\gamma_1=\gamma_2=\gamma$ and $\eta_1=\eta_2$. We may think of $\gamma<0$ as the typical solar 
case in which transport inwards by convective overshoot is expected to dominate the buoyant rise of magnetic field. 

\item{\bf Dominant transport in the convection zone}\\
Here we assume that the net transport is much stronger in $z>0$ and that effects due to magnetic buoyancy in $z>0$ are small 
in comparison with convective transport. We therefore consider $\gamma_2=0$ and investigate the effects of increasing the 
magnitude of $\gamma_1<0$. This is motivated by what is believed to be the case at the base of the solar convection zone. We 
compare the case $\eta_2=\eta_1$ with the more relevant case $\eta_2 \ll \eta_1$.
\end{itemize}

For chosen values of the $\gamma$-effect and diffusion coefficients, the dispersion relation is solved to yield a critical dynamo 
number $D_c$ and corresponding critical frequency $\omega_c$ as functions of the wavenumber. Dynamo 
action sets in for $D>D^* \equiv \textrm{min}_k \left[D_c(k)\right]$, with preferred wavenumber $k=k^*$ and 
corresponding frequency $\omega_c(k^*)$. We shall say that transport hinders dynamo action if 
increasing the magnitude of $\gamma$ increases the critical dynamo number $D^*$, while if $D^*$ is somewhere a decreasing 
function of $\gamma$ then transport aids the dynamo instability. The effect of transport on the period of the dynamo wave ($T 
= 2 \pi/\omega_c$), reflecting the time taken to reverse the sign of the magnetic field, will also be investigated, as will its role in 
modifying the spatial profile of the magnetic field. 

Finally, it is also of interest to consider the effect of transport on the phase relation between the radial and 
azimuthal fields. Observations indicate that $B_r B_{\phi}<0$ at active latitudes, i.e.\ the two components have a phase shift 
$\phi=\pi$ (see, for example, Stix 1976; Yoshimura 1976; Schlichenmaier \& Stix 1995). This phase difference has traditionally 
been difficult to reproduce with the simplest $\alpha\omega$-dynamo models that rely upon diffusive communication: given the 
sign of $\partial \Omega/\partial r$ from helioseismology, the sign of $\alpha$ required for equatorward migration { ($\alpha<0$ 
in the northern hemisphere, $\alpha>0$ in the southern)} generates poloidal field whose sign is inconsistent with the above 
phase relation (see Parker 1987; Sch\"ussler 2005). It is therefore of interest to investigate whether more complex 
$\alpha\omega$-models can reproduce the observed behaviour. We note that a non-zero phase difference is inherent in the 
dynamo equations, even for models with spatially coincident generation regions (see Parker 1955, for example). For our 
dynamo models, the phase difference will be found by examination of the temporal profiles of the azimuthal and radial fields. 
For example, at $x=z=0$ we find 
\begin{equation}
\Re\left[B(t)\right] \sim \cos\left(\frac{2 \pi t}{T}\right), \qquad \Re\left[\frac{\partial A(t)}{\partial x}\right]\sim 
\cos\left(\frac{2 \pi t}{T}+\phi\right) ,
\end{equation}
where $\phi$ is the phase difference. It is important also to note that the observational evidence for the phase lag arises from the 
behaviour of the two components of field at the solar surface. Since the toroidal field is created in the tachocline at 
approximately $0.7R_{\odot}$ and is subject to complex processes operating throughout the bulk of the convection zone 
during its rise to the surface, the phase lag is likely to vary with depth. In this investigation we choose to evaluate $\phi$ at the 
interface of the convection zone and tachocline ($z=0$). For the model of infinite extent in $z$ (Model~I) this is done owing to 
the lack of any other preferred location, whereas in the vertically bounded model (Model~II) this is done to minimise the role of 
boundary conditions. 

Before turning to the details of the models, it is interesting to note a common characteristic of both models. As can be 
observed directly from equations (\ref{eq:dynamo_z>0}) and (\ref{eq:dynamo_z<0}), in the large wavenumber 
limit the evolution is governed by the diffusive term. Thus $D_c \rightarrow \infty$ as $k \rightarrow \infty$. 
Furthermore, since in this limit the critical dynamo number and corresponding frequency are independent of the magnitude of 
the transport mechanism, all the stability curves for different magnitudes of $\gamma_{1}$ and $\gamma_2$ collapse onto that 
for $\gamma_1=\gamma_2=0$. 


\section{Model~I: Spatially infinite domain}
\label{sec:Model1}

In this section we solve the dynamo equations (\ref{eq:dynamo_z>0}) and (\ref{eq:dynamo_z<0}) in a domain that is 
unbounded in the $z$-direction (see Figure~\ref{fig:geometry}a), with boundary conditions 
\begin{equation}
\label{eq:bconds}
a(z) \rightarrow 0, \quad b(z) \rightarrow 0 \quad \textrm{ as } |z| \rightarrow \infty,
\end{equation}
since neither the shear nor the $\alpha$-effect alone can sustain dynamo action. Following Parker (1993), we seek 
solutions of the form (\ref{eq:solns}) with
\begin{equation}
\label{eq:B1A1}
b_{1}(z)=C\exp{[-(S+\textrm{i}Q)z]}, \qquad a_{1}(z)=(F+Ez)\exp{[-(S+\textrm{i}Q)z]} \quad \textrm{ in } z>0,
\end{equation}
and
\begin{equation}
\label{eq:A2B2}
a_{2}(z)=J \exp{[(s+\textrm{i}q)z]}, \qquad b_{2}(z)=(L+Mz)\exp{[(s+\textrm{i}q)z]} \quad \textrm{ in } z<0.
\end{equation}
Here $C$, $S$, $Q$, $s$, $q$, $\sigma$, $\omega,k$ are real constants, while we allow $E$, $F$, $J$, $L$, $M$ to be 
complex. Since our analysis is linear, there exists one arbitrary amplitude, which we take as $C$. The boundary conditions 
(\ref{eq:bconds}) require that $S>0$ and $s>0$. Applying continuity and the jump relations 
(\ref{eq:jump}) then leads to the relation (see Appendix~A)
\begin{eqnarray}
\label{eq:m2disp}
&&\left[S+\textrm{i}Q+s+\textrm{i}q\right]\left[(S+\textrm{i}Q)+\mu^2(s+\textrm{i}q)
+ \frac{(\gamma_{1}-\gamma_{2})}{\eta_{1}}\right]
\left[2(S+\textrm{i}Q)
+\frac{\gamma_{1}}{\eta_{1}}\right]\left[2(s+\textrm{i}q)-\frac{\gamma_{2}}{\eta_{2}}\right]=
\textrm{i}k^4 \left(\frac{\alpha G}{\eta_1^2 k^3}\right),
\end{eqnarray}
where 
\begin{equation}
\label{eq:sigom}
S+\textrm{i}Q=-\frac{\gamma_1}{2 \eta_1} \pm \frac{1}{2 \eta_1} \sqrt{\gamma_1^2+4 \eta_1 
(\sigma+\textrm{i} \omega + \eta_1 k^2)}, \qquad s+\textrm{i}q=\frac{\gamma_2}{2 \eta_2} \pm \frac{1}{2 \eta_2} 
\sqrt{\gamma_2^2+4 \eta_2 (\sigma+\textrm{i} \omega + \eta_2 k^2)},
\end{equation}
and $\mu^2=\eta_{2}/\eta_{1}$. In order to satisfy the conditions at infinity, the signs in expressions (\ref{eq:sigom}) are to be 
chosen so that $S>0$ and $s>0$. The term in parentheses on the right-hand side of equation (\ref{eq:m2disp}) is a 
dimensionless measure of the relative strength of induction to diffusion and is defined as the dynamo number for this problem, 
i.e.\ $D \equiv \alpha G /\eta_{1}^2 k^3$. Without loss of generality, we take $D>0$. There exist similar solutions for $D<0$ 
but they propagate in the opposite direction.  Note that, owing to the lack of any physical length scale in this problem, the 
dynamo number is here defined in terms of the wavenumber $k$ (e.g.\ Parker 1955). Thus when we consider the effects of 
transport on the onset of the dynamo instability as a function of wavenumber we consider the dependence of $\gamma$ on 
$k^3 D_c \equiv \left(\alpha G/\eta_1^2\right)_c$. 

We note that in the absence of the $\gamma$-effect the model reduces to that of Parker (1993). In the case of uniform 
diffusion, ($\eta_1=\eta_2$) equations (\ref{eq:m2disp}) and (\ref{eq:sigom}) then yield
\begin{equation}
\label{eq:parker_mu1}
\left(\frac{\alpha G}{\eta_1^2}\right)_c=\pm 32 k^3, \qquad \omega_c =\pm \eta_1 k^2.
\end{equation} 
In what follows we take the $(\alpha G/\eta_1^2)_c>0$, $\omega_c>0$ solutions as our starting point when considering the 
effects of non-zero transport; i.e., as the magnitude of $\gamma$ increases, our solutions are smoothly linked to expressions 
(\ref{eq:parker_mu1}) with the positive signs.  As discussed earlier, we split the general case into the following subcases.

\subsection{Bidirectional transport and uniform diffusion}
\label{sec:m2_bidirectional}

\begin{figure}
\begin{center}
\resizebox{0.95\textwidth}{!}{ \includegraphics{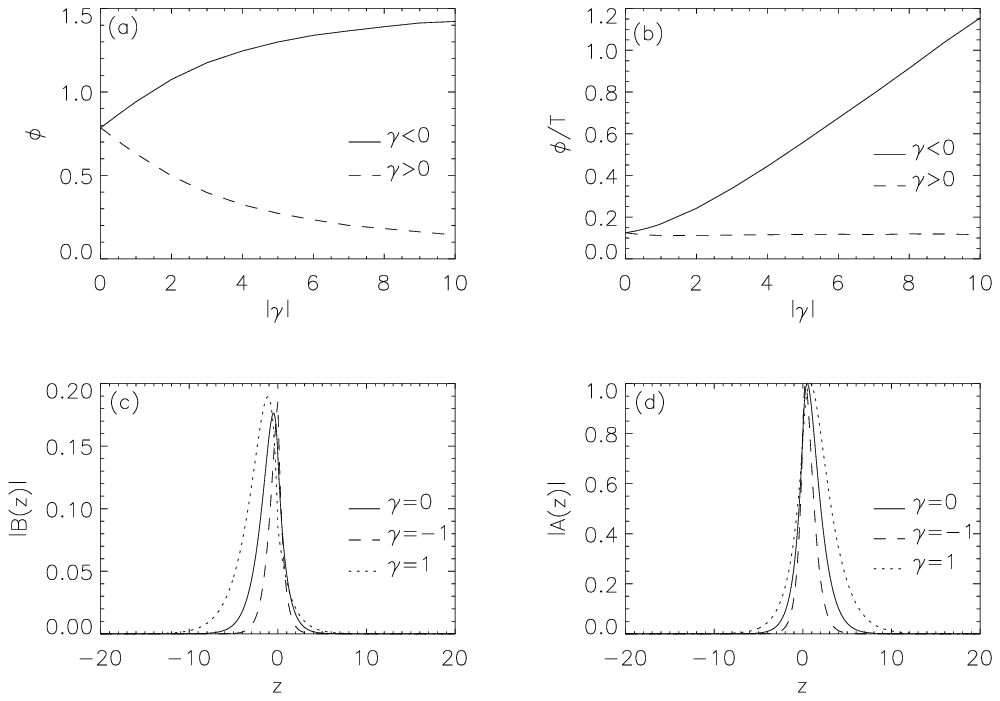}}
\end{center}
\caption{Model~I: Bidirectional transport and uniform diffusion ($\gamma_{1}=-\gamma_2=\gamma$, 
$\eta_{2}=\eta_{1}=1$). (a) Phase difference $\phi$, (b) Phase difference divided by the period $\phi/T$, (c) Toroidal field, 
(d) Poloidal field. 
$\gamma=0$ represents no transport, $\gamma=-1$ transport towards the interface and $\gamma=1$ 
transport away from the interface. The eigenfunctions are for $k=1$ and the phase difference is calculated 
for $k=1$ and $x=z=0$.} 
\label{fig:Parker_g_up_down}
\end{figure}

We begin with the case $\gamma_1=-\gamma_2=\gamma$ and $\eta_1=\eta_2=\eta$. Relations 
(\ref{eq:m2disp}) and (\ref{eq:sigom}) then yield
\begin{equation}
\left(\frac{\alpha G}{\eta_1^2}\right)_{c}=\frac{2}{\eta^3}(\gamma^2+8\eta^2k^2)(\gamma^2+4\eta^2k^2)^{1/2},\qquad 
\omega_c=\frac{k}{2}(\gamma^2+4\eta^2k^2)^{1/2}
\end{equation}
It is clear that increasing $\gamma$ increases both $\left(\alpha G/\eta_1^2\right)_c$ and $\omega_c$. Thus transport hinders 
the onset of the dynamo instability and decreases the time taken to reverse the field polarity. Furthermore, both relations are 
independent of the sign of $\gamma$.  Thus, although one may expect the dynamo properties of the 
two cases $\gamma<0$ and $\gamma>0$, representing flux transport towards and away from the interface in both 
domains, respectively, to be very different, the critical parameters governing the onset of dynamo action are not affected by the 
direction of flux transport. We believe this unusual property to be a consequence of the infinite extent of the model.

It is noted that the parameters governing the $z$-structure of the solution are sensitive to the sign of the $\gamma$-effect, {\it 
viz.} $S+\textrm{i}Q=s+\textrm{i}q=-\gamma / 2 \eta +\sqrt{\gamma^2+4 \eta (\sigma+\textrm{i} \omega+\eta k^2 )}/2 \eta$. 
As a 
consequence, the dependence on $\gamma$ of the phase difference between the azimuthal and radial fields is different for 
positive and negative $\gamma$, as shown in Figures~\ref{fig:Parker_g_up_down}(a,b). In the case when transport is directed 
towards the interface ($\gamma<0$), the phase lag is increased as $|\gamma|$ increases, while transport away from the interface 
($\gamma>0$) brings the two components of field more in phase. 

The structure of the azimuthal and radial fields is illustrated in Figures~\ref{fig:Parker_g_up_down}(c,d). As expected, the 
eigenfunctions are narrower (wider) when transport is towards (away from) the interface. 

\subsection{Unidirectional transport and uniform diffusion}
\label{sec:Parker_eqge}

Given that flux transport towards the interface hinders the excitation of the dynamo instability in exactly the same manner as 
does transport away from the interface, it appears intuitively unlikely that a unidirectional 
$\gamma$-effect operating throughout the domain will enhance the efficiency of dynamo action. Indeed, in the case 
$\gamma_1=\gamma_2=\gamma$ and $\eta_1=\eta_2=\eta$, analytic investigations yield
\begin{equation}
\label{eq:Dcparkereqneqg}
\left(\frac{\alpha G}{\eta_1^2}\right)_{c}=32 k^3\Big(1+\frac{\gamma^2}{4k^2\eta^2}\Big)^2, \qquad \omega_c=
\eta k^2 \left(1+\frac{\gamma^2}{4k^2\eta^2}\right).
\end{equation}
The dynamo is most easily excited at $k=k_c=|\gamma|/(2\eta\sqrt{3})$, where $\left(\alpha G/\eta_1^2\right)_c=
64 |\gamma|^3/\left(3\eta^3\sqrt{3}\right)$. Thus transport again hinders the onset of the dynamo instability. 
The frequency increases with increasing magnitude of the transport, and the dynamo prefers to generate 
waves with ever decreasing wavelength ($\lambda=2\pi/k$).

It may also be shown analytically that at $z=0$ increasing the magnitude or changing the direction of the transport 
has no effect on the phase difference between the toroidal and radial fields, as $\phi=\pi/4$ always. 
Furthermore, since the phase relation remains constant even though the critical parameter governing the onset of 
dynamo action can increase dramatically, the phase lag is not a property that controls the efficiency of dynamo action 
in the regime studied here. 
 
\begin{figure}
\begin{center}
\resizebox{0.95\textwidth}{!}{\includegraphics{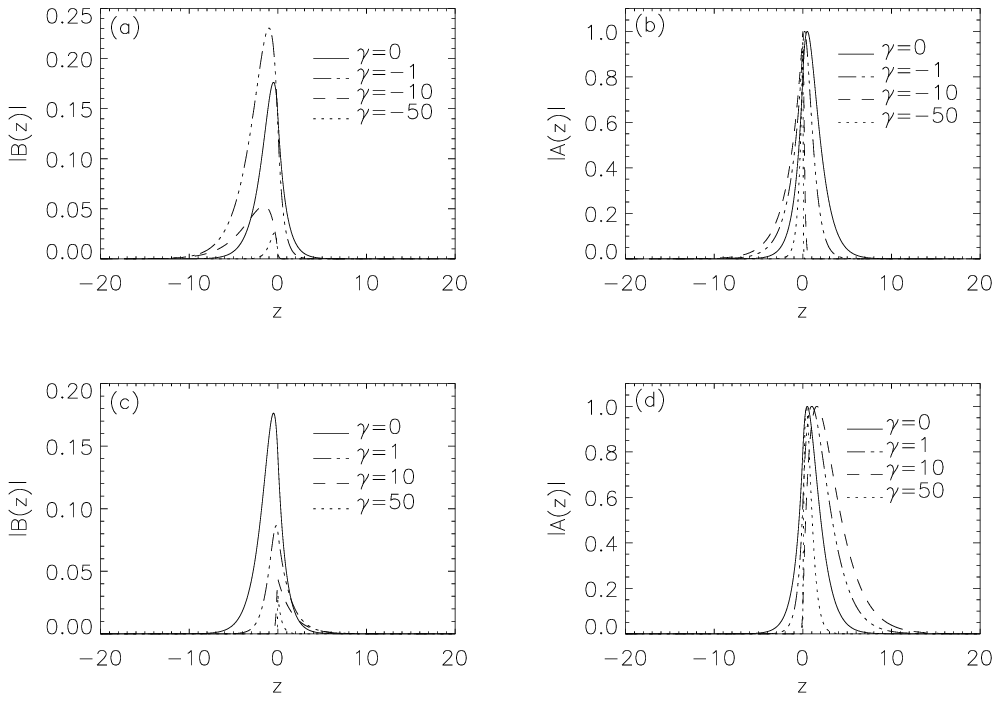}}
\end{center}
\caption{Model~I: Unidirectional transport and uniform diffusion ($\gamma_1=\gamma_2=\gamma$, $\eta_1=\eta_2=1$,
$k=1$). (a,b) Toroidal and poloidal fields when the $\gamma$-effect is
directed downwards. (c,d) As for (a,b) with the $\gamma$-effect directed upwards.  For $\gamma=-50,10,50$,
the azimuthal field has been multiplied by $10,50,1000$, respectively.}
\label{fig:m2_efns}
\end{figure}

Figure~\ref{fig:m2_efns} compares the structure of the azimuthal and radial fields for unidirectional transport
directed downwards with that for upwards transport. Downwards transport reduces the decay length of the fields in 
$z>0$, while upwards transport reduces the decay length in $z<0$. 

\subsection{Dominant transport effect in $z>0$}

We now consider the effects of removing the $\gamma$-effect from $z<0$. In these cases the toroidal field generated in the 
shear layer will rely upon diffusion to transport it into $z>0$, where it will act as a source for poloidal field regeneration.

\subsubsection{Uniform diffusion}
\label{sec:m2_mu1g20}

\begin{figure}
\begin{center}
\resizebox{0.95\textwidth}{!}{ \includegraphics{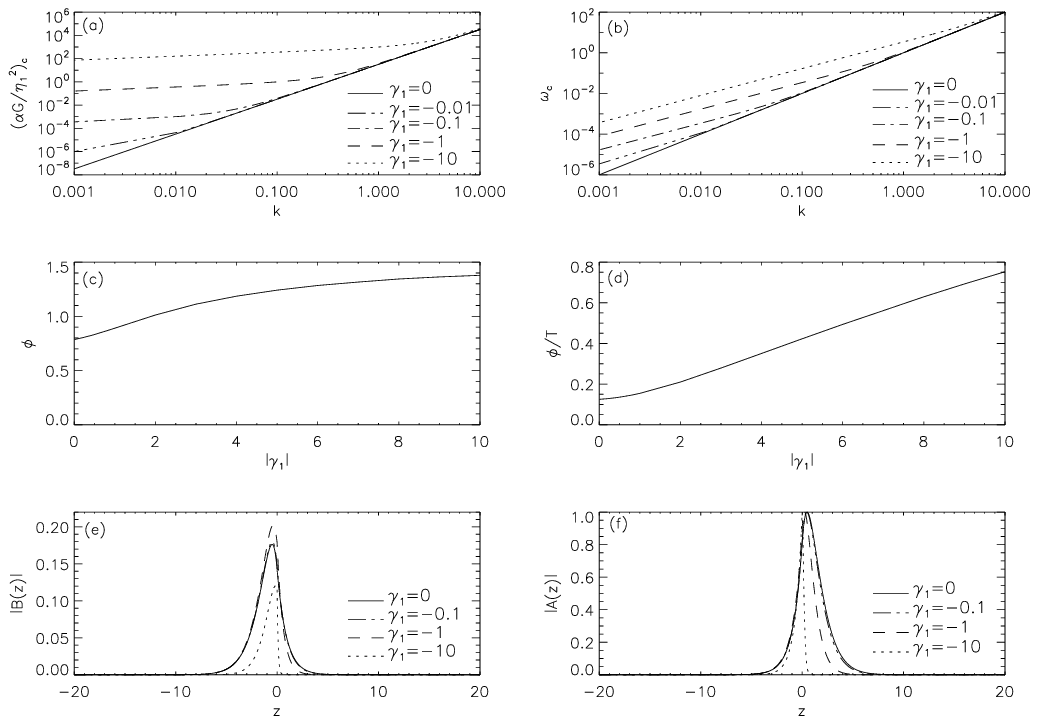}}
\end{center}
\caption{Model~I: Dominant transport in $z>0$ and uniform diffusion ($\gamma_{1}<0$, $\gamma_2=0$, 
$\eta_{2}=\eta_{1}=1$). (a) Critical parameter \(k^3D_{c}\), (b) Critical frequency $\omega_c$, 
(a) Phase difference $\phi$, (b) Phase difference divided by the period $\phi/T$, (c) Toroidal field, (d) Poloidal field. 
The eigenfunctions are for $k=1$ and the phase difference is calculated 
for $k=1$ and $x=z=0$.}
\label{fig:m2_g20_mu1}
\end{figure}

Figures~\ref{fig:m2_g20_mu1}(a,b) illustrate the marginal stability curves for the case $\gamma_1<0$, $\gamma_2=0$ 
and $\eta_1=\eta_2=\eta=1$. For small wavenumbers
\begin{equation}
\left(\frac{\alpha G}{\eta_1^2}\right)_c \approx \left[\frac{4 \gamma_1^8}{\eta^8}\right]^{1/3} k^{1/3}, \qquad 
\omega_c \approx \left[\frac{\eta \gamma_1^2}{2}\right]^{1/3} k^{4/3}, \quad \textrm{as } k \rightarrow 0,
\end{equation}
which may be compared with (\ref{eq:parker_mu1}) for the case of $\gamma=0$. Thus, again, transport hinders 
the onset of dynamo action and increases the frequency of the dynamo waves. It is interesting to note that in comparison with 
the uniform transport case of \S\ref{sec:Parker_eqge}, the removal of the $\gamma$-effect from $z<0$ results in a phase 
difference that is dependent on the magnitude of $\gamma_1$ [Figure~\ref{fig:m2_g20_mu1}(c)]. In this case increasing 
$\gamma_1$ increases the phase difference. The ratio of the phase difference to the period of the wave also increases 
[Figure~\ref{fig:m2_g20_mu1}(d)]. The eigenfunctions, shown in Figures~\ref{fig:m2_g20_mu1}(e,f), exhibit similar behaviour 
to the case of a unidirectional transport, with the fields being distorted towards the interface from above (the direction in which 
the transport operates).

\subsubsection{Reduced diffusion in $z<0$}

\begin{figure}
\begin{center}
\resizebox{0.95\textwidth}{!}{ \includegraphics{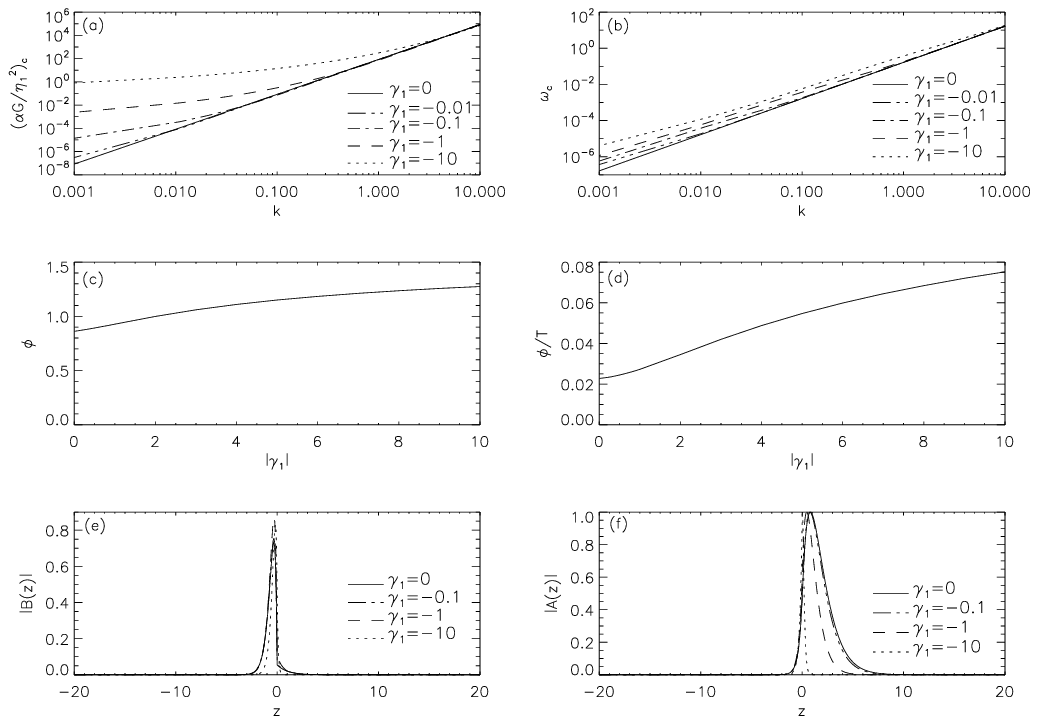}}
\end{center}
\caption{ As for Figure~\ref{fig:m2_g20_mu1} with $\gamma_1<0$, $\gamma_2=0$, $\eta_1=1$, $\eta_2=0.01$. 
}\label{fig:m2_g20_mus}
\end{figure}

The marginal stability curves for the case $\gamma_1<0$, $\gamma_2=0$, $\eta_1=1$ and $\eta_2=0.01$ are 
shown in Figures~\ref{fig:m2_g20_mus}(a,b). Again transport hinders the operation of the dynamo and 
decreases the period of the dynamo wave. The phase difference between the radial and azimuthal 
fields behaves in a qualitatively similar manner to the cases investigated above in which the $\gamma$-effect is directed 
downwards, with both $\phi$ and $\phi/T$ increasing as the magnitude of $\gamma_1$ increases 
[Figures~\ref{fig:m2_g20_mus}(c,d)]. The behaviour of the eigenfunctions appears to be such that the downwards-directed 
transport reduces the decay lengths in $z>0$ [Figures~\ref{fig:m2_g20_mus}(e,f)]. 
 
Comparing these results with the uniform diffusion case described above (\S\ref{sec:m2_mu1g20}), one finds that in the 
absence of $\gamma_1$ the dynamo is easier to excite in the case of uniform diffusion than for $\eta_2/\eta_1=0.01$, 
for those wavenumbers considered. Indeed, in the limit $\mu^2=\eta_2/\eta_1 \rightarrow 0$, Parker (1993), in the absence of 
transport, derived the relations 
\begin{equation}
\label{eq:m2_g0smu_dcwc}
\left(\frac{\alpha G}{\eta_1^2}\right)_c \approx \frac{2^{5/3}}{\mu^{4/3}}k^3, \qquad \omega_{c} \approx 
\frac{\eta_{1} \mu^{2/3}}{2^{1/3}}k^2,
\end{equation}
while for $\mu^2=1$ we have expressions (\ref{eq:parker_mu1}). Clearly, in the absence of non-diffusive transport, 
$\mu^2 \rightarrow 0$ yields a greater critical dynamo number and a smaller frequency than for the case of uniform 
diffusion. However, our results reveal that for finite $\gamma_1$ the situation is reversed at a certain wavenumber, 
$k=\tilde k(\gamma_1)$, say. For wavenumbers less than $\tilde k$ the dynamo is easier to excite in the case 
with smaller $\eta_2$, while for $k>\tilde k$ dynamo action occurs for smaller values of $(\alpha G/\eta_1^2)$ in 
the case with uniform diffusion. This can be seen by comparing Figures \ref{fig:m2_g20_mu1}(a) and 
\ref{fig:m2_g20_mus}(a) for the case $\gamma_1=-1$, where $\tilde k(-1) \approx 0.3$. For those $\gamma_1$ considered, 
the frequencies are larger in the case of uniform 
diffusion than they are for $\eta_2=0.01$. However, it should be noted that for different spatial profiles of $\gamma$ 
it is not always the case that uniform diffusion results in a larger frequency (for example, compare the frequencies in 
Figure~\ref{fig:m2_eqg_eta2}(b) for $\eta_2=1$ and $\eta_2=0.01$ in the case of 
$\gamma_1=\gamma_2=-1$). Finally, inspection of the eigenfunctions shows that the reduced $\eta_2$ results in a smaller 
vertical 
length scale for the toroidal field in $z<0$, as expected (Parker 1993; Tobias 1996). 

\subsection{The competition between the $\gamma$-effect and diffusion}
\label{sec:gamma_diffn}

\begin{figure}
\begin{center}
\resizebox{0.95\textwidth}{!}{ \includegraphics{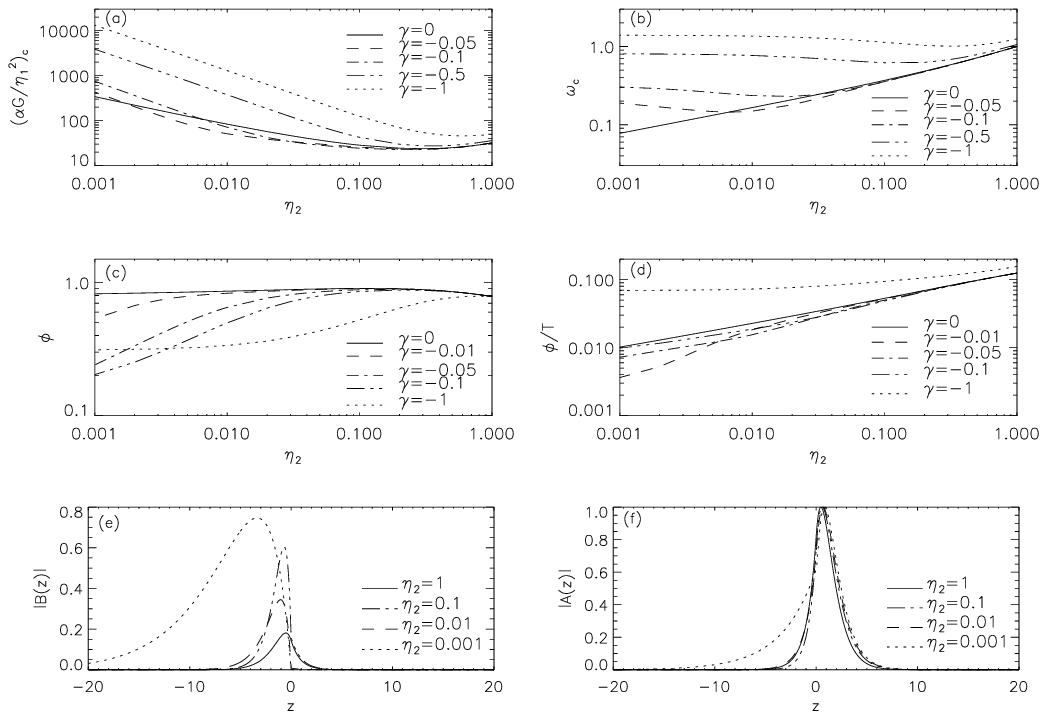}}
\end{center}
\caption{ Model~I: The competition between the $\gamma$-effect and diffusion ($\gamma_{1}=\gamma_{2}=\gamma<0$, 
$\eta_{1}=1$, $k=1$). (a) Critical parameter $k^3 D_c$, (b) Critical frequency $\omega_c$, (c) Phase difference 
$\phi$, (d) Phase difference divided by the period $\phi/T$, (e) Toroidal field, (f) Poloidal field. The eigenfunctions are for
$\gamma=-0.05$; for $\eta_2=0.01$ and $\eta_2=0.001$ the toroidal field has been divided by $10$ and $20$, respectively. 
The phase difference is calculated for $x=z=0$.}\label{fig:m2_eqg_eta2}
\end{figure}

It is instructive to consider the effect of varying the diffusivity in $z<0$ in the absence of the $\gamma$-effect and to compare 
the results with the corresponding situation when $\gamma$ is non-zero. We consider a fixed value 
of the wavenumber, here chosen to be $k=1$, and we take $\gamma_1=\gamma_2=\gamma$ for simplicity. 
Figure~\ref{fig:m2_eqg_eta2}(a) illustrates that in both cases the effect on $(\alpha G/\eta_1^2)_c$ of varying $\eta_2$ is 
non-monotonic. Decreasing $\eta_2$ from unity initially allows a more effective production of toroidal field in $z<0$ and thus 
the dynamo becomes easier to excite. However, there exists a transition point 
($\eta_2^*(\gamma)$, say) after which decreasing the diffusion hampers the coupling between the two regions sufficiently that 
the onset of dynamo action is delayed.

Figure~\ref{fig:m2_eqg_eta2}(b) illustrates that when the $\gamma$-effect is absent the frequency decreases 
monotonically as $\eta_2$ decreases, while when $\gamma \ne 0$ there are two distinct regimes. The first is when 
$\gamma \ll k \eta_2$, in which the frequency of the wave is governed by diffusive transport between the layers. 
Here the frequency decreases as $\eta_2$ decreases. The second regime begins when $k \eta_2$ becomes 
comparable with $\gamma$, and the frequency is then controlled both by diffusion and the $\gamma$-effect. 
As $\eta_2$ is decreased further there is an increase in the frequency of the dynamo wave until $\omega_c$ 
eventually saturates at a constant value. In the final state the frequency of the wave is governed by the $\gamma$-effect.

Figure~\ref{fig:m2_eqg_eta2}(c) shows that the phase difference between the toroidal and radial fields is also governed 
both by diffusion and the $\gamma$-effect. The phase difference is typically a non-monotonic function of the diffusion, 
even for the case $\gamma=0$. The two components of field initially become more out of phase as the reduced diffusivity 
favours the generation of toroidal field in $z<0$. However, as the diffusion is reduced further the phase difference decreases.
 The ratio of the phase difference to the period of the wave is shown in Figure~\ref{fig:m2_eqg_eta2}(d).

The effect of reducing the diffusivity in $z<0$ on the spatial structure of the fields is illustrated in 
Figures~\ref{fig:m2_eqg_eta2}(e,f). In the case $\gamma=0$ (compare Figure~\ref{fig:m2_efns}(a) with 
\ref{fig:m2_g20_mus}(e)) the effect of decreasing $\eta_2$ is increasingly to confine the toroidal field beneath 
$z=0$, as first shown by Parker (1993), while when $\gamma \ne 0$ a sufficiently small diffusion in $z<0$ allows 
significant amounts of field to be advected away from the interface.

\section{Model~II: Spatially confined domain}
\label{section:Parkerbdd}

In this section we investigate a more realistic model whose vertical extent is limited to $z \in [-L,L]$, where $L$ is finite (see 
Figure~\ref{fig:geometry}b). All other aspects of the 
geometry remain as in Model~I. One consequence of the finite domain is that we may meaningfully non-dimensionalise the 
system. We let
\begin{equation}
[\mathbf{x},\alpha,G=\frac{du}{dz},\eta,\gamma,B,A,t]\equiv[L \mathbf{x}',\alpha _0\alpha',G_0G',\eta _0 \eta ',
\gamma _0 \gamma ',B_0B',\frac{\alpha _0 L^2 B_0}{\eta _0}A', \frac{L^2}{\eta_0}t'],
\end{equation} 
where primed variables are dimensionless. The dynamo equations then become (dropping the primes) 
\begin{equation}
\label{eq:nondima} 
\frac{{\partial}{A_1}}{{\partial}{t}} = {\alpha B
+\eta_1 \nabla^{2} A_1 - Y_1 \frac{{\partial}{A_1}}{{\partial}{z}}}, \qquad 
\frac{{\partial}{B_1}}{{\partial}{t}} =
{\eta_1\nabla^{2}B_1-Y_1 \frac{{\partial}{B_1}}{{\partial}{z}}} \qquad 
\textrm{ in } z>0,
\end{equation}
and 
\begin{equation}
\label{eq:nondimb} 
\frac{{\partial}{A_2}}{{\partial}{t}} = \eta_2 \nabla^{2} A_2- Y_2 \frac{{\partial}{A_2}}{{\partial}{z}}, \qquad 
\frac{{\partial}{B_2}}{{\partial}{t}} =
{\eta_2\nabla^{2}B_2+DG\frac{{\partial}{A_2}}{{\partial}{x}}-Y_2\frac{{\partial}{B_2}}{{\partial}{z}}} \qquad 
\textrm{ in } z<0.
\end{equation}
Here $Y_{1,2}= R_{\gamma}\gamma_{1,2}$ is a dimensionless measure of the $\gamma$-effect and 
$R_{\gamma}=\gamma_0 L/\eta_0$ is its Reynolds number. The dynamo number is defined as 
$D \equiv \alpha_0 G_0 L^3/{\eta_0}^2>0$. 

Travelling wave solutions of the form (\ref{eq:solns}) are sought in each of $z>0$ and $z<0$ and the solutions are 
matched at the interface $z=0$ by the jump relations (\ref{eq:jump}). The boundary conditions that approximate the 
matching onto an external potential field (see Zeldovich, Ruzmaikin \& Sokoloff 1983) are,
\begin{equation}
\label{eq:boundary1ab}
B_1(x,z=1,t)=0, \qquad \frac{\partial{A_1}}{\partial z}(x,z=1,t)=0,
\end{equation} 
\begin{equation}
\label{eq:boundary2ab}
B_2(x,z=-1,t)=0, \qquad \frac{\partial{A_2}}{\partial z}(x,z=-1,t)=0.
\end{equation}
The resulting dispersion relation, along with details of its derivation, can be found in Appendix~B. Again we split the general 
case into a number of subcases.

\subsection{Bidirectional transport and uniform diffusion}

\begin{figure}
\begin{center}
\resizebox{0.95\textwidth}{!}{\includegraphics{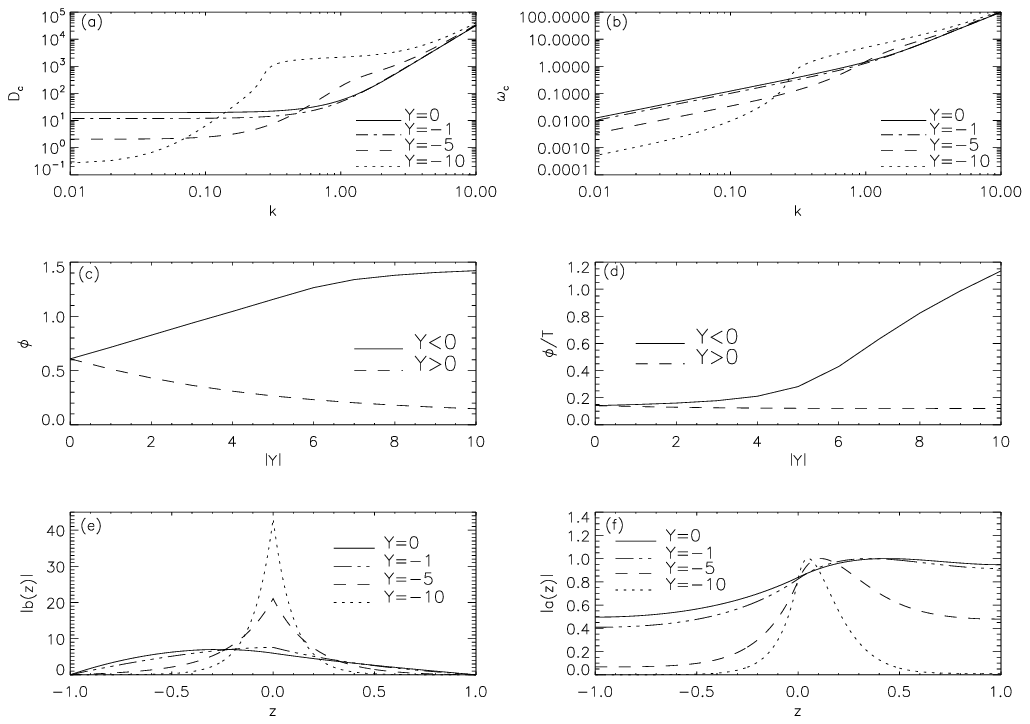}}
\end{center}
\caption{Model~II: Bidirectional transport with uniform diffusion ($Y_1=-Y_2=Y$, $Y<0$, $\eta_{1}=\eta_{2}=1$). (a) 
Critical dynamo number $D_c$, (b) Critical frequency $\omega_c$, (c) Phase difference $\phi$, (d) 
Phase difference divided by the period $\phi/T$, (e) Toroidal field, (f) Poloidal field. The toroidal field has been
divided by factors of 2 and 4 for $Y=-5$ and $Y=-10$ respectively ($k=1$). The phase difference and $\phi/T$ are compared 
with the case of $Y>0$ representing transport away from the interface in both $z>0$ and $z<0$ 
($k=1$, $x=z=0$). }\label{fig:m3_gud_mu1}
\end{figure}

It is instructive first to investigate the case in which the $\gamma$-effect is directed towards the interface from both above and 
below. We therefore take $Y_1=-Y_2=Y$ with $Y<0$ and for simplicity we consider the case of uniform diffusion, 
$\eta_1=\eta_2=\eta$. The marginal stability curves and corresponding frequency profiles for different magnitudes of the
transport are shown in Figures~\ref{fig:m3_gud_mu1}(a,b). For $k \gg 1$, all profiles collapse onto 
\begin{equation}
\label{eq:m3_eqgeqn_lk}
D_c \approx 32 \left(\frac{\eta^2 k^3}{\alpha G}\right), \qquad \omega_c \approx \eta k^2,
\end{equation}
as the system reduces to Model~I in the large wavenumber limit (cf. (\ref{eq:parker_mu1}); Parker 1993). The 
numerical results illustrate that the dynamo instability is easiest to excite at $k \ll 1$, where
$D_{c} \approx D_{0}(Y, \eta)$ and $\omega_{c} \approx \omega_{1}(Y, \eta)k.$
Analytic investigations for $Y=0$ yield $D_{0}(0,\eta)=16 \eta \omega_{1}$ and $\omega_1(0,\eta)=\eta \sqrt{3/2}$.
Figures~\ref{fig:m3_gud_mu1}(a,b) illustrate that increasing the magnitude of the transport decreases both $D_0$ and the 
corresponding frequency of the dynamo wave. Thus here the transport effect {\it aids} the onset of the dynamo instability. The 
time taken for the field to reverse sign increases. 

It is interesting that both of the above results concerning the dynamo number and frequency are the direct opposite of any 
situation that we have previously encountered, i.e.\ the critical dynamo number and frequency in \S\ref{sec:Model1} were 
increasing functions of $\gamma$. In particular, this is true for the corresponding case of Model~I 
(see \S\ref{sec:m2_bidirectional}), and we are led to the conclusion that boundaries in the $z$-direction play a crucial role 
when considering the effects of transport on the efficiency of dynamo action. In fact, investigating the complementary situation 
in which $\gamma$ is directed away from the interface in both domains ($Y_1=-Y_2=Y$, $Y>0$) yields numerical results with 
coefficients $D_0$ and $\omega_1$ monotonically increasing functions of $Y$. Recalling that Model~I was dependent only on 
the magnitude of the bidirectional transport mechanism, we conclude that the boundaries in the $z$-direction cause the model 
to become sensitive to the direction of flux transport, which is more realistic.

Figures~\ref{fig:m3_gud_mu1}(c,d) illustrate the effect of transport on the phase difference between the azimuthal and radial 
fields, and on the ratio of the phase difference and the period of the wave. For comparison, also presented is the case when 
transport acts away from the interface in both regions. Transporting flux towards the interface causes the two components of 
field to become more out of phase whereas when $Y$ is directed away from the interface the phase difference decreases, in 
agreement with Model~I.

Conveying magnetic flux towards the interface favours field generation there, causing both the toroidal and poloidal fields to 
become increasingly localised around $z=0$ [Figures~\ref{fig:m3_gud_mu1}(e,f)]. In the contrasting situation of flux transport 
away from the interface ($Y>0$) the main effect is to cause the peak in toroidal magnetic field to move towards the lower 
boundary (behaviour that is qualitatively similar to that shown in 
Figure~\ref{fig:m3_eqg_mu1}(e) for uniform downwards transport), whereas there is little change in the $z$-structure of the 
poloidal field.

\subsection{Unidirectional transport and uniform diffusion}
\label{section:Parkerbddeqneqg}

\begin{figure}
\begin{center}
\resizebox{0.95\textwidth}{!}{ \includegraphics{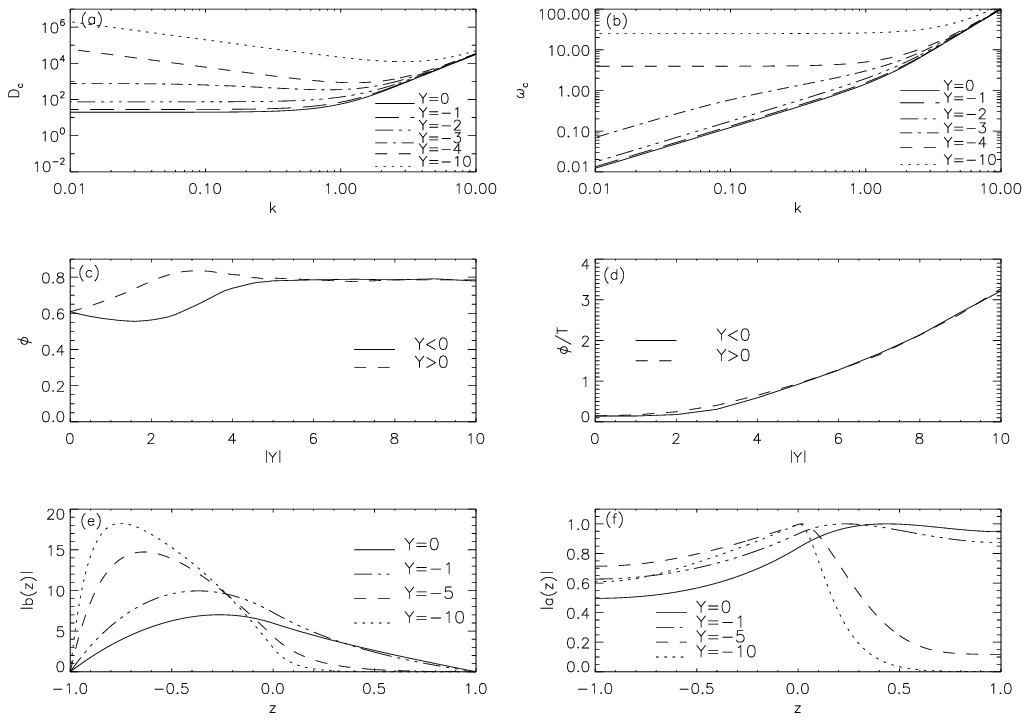}}
\end{center}
\caption{As for Figure~\ref{fig:m3_gud_mu1} with $Y_{1}=Y_{2}=Y<0$, $\eta_{1}=\eta_{2}=1$. The toroidal field has 
been divided by factors of 10 and 50 for $Y=-5$ and $Y=-10$ respectively.} \label{fig:m3_eqg_mu1}
\end{figure}

When the transport velocity is uniform throughout the whole domain ($Y_1=Y_2=Y$) and the diffusivities are equal, the  
dispersion relation is independent of the sign of $Y$. Figures~\ref{fig:m3_eqg_mu1}(a,b) illustrate that for $k \ll 1$ the critical 
parameters exhibit different behaviours depending upon the magnitude of the transport: for $|Y|<Y_{s}(\eta)$, say, the most 
readily destabilised mode satisfies $D_c \approx D_0(Y, \eta)$, 
$\omega_c \approx \omega_1(Y,\eta)k$, whereas for $|Y| >Y_{s}$ we have
$D_{c} \approx D_{-1}(Y,\eta)k^{-1}$, $\omega_{c} \approx \omega_{0}(Y,\eta)$ [for $\eta=1$, $Y_{s} \approx 3.1$ ].
Since at large wavenumbers the marginal stability profiles must approach the results (\ref{eq:m3_eqgeqn_lk}) for 
Model~I, it follows that for those $|Y| >Y_{s}$ there exists a preferred wavelength for the dynamo instability 
(for example, $\lambda^* \approx 2 \pi/3$ for $Y=-10$).  Figures~\ref{fig:m3_eqg_mu1}(a,b) illustrate that transport 
hinders the excitation of the dynamo and decreases the period of the wave.

The phase difference between the azimuthal and radial fields follows a non-monotonic behaviour as the magnitude of $Y$ 
is increased, and is dependent upon the direction of flux transport [Figures~\ref{fig:m3_eqg_mu1}(c,d)]. When $Y<0$ the two 
fields are most in phase for $Y \approx -1.6$, which can be compared with the situation for $Y>0$ where the stationary point 
is a maximum and the fields are most out of phase for $Y \approx 3.2$.
 
The spatial structure of the toroidal and poloidal fields in the presence of a uniform transport velocity directed towards the 
boundary $z=-L$ is shown in Figures~\ref{fig:m3_eqg_mu1}(e,f). Interestingly, increasing the magnitude of the transport 
results in the majority of the poloidal field residing in $z<0$, even though it is generated in $z>0$.

\subsection{Dominant transport effect in $z>0$}

We now consider the case motivated by the Sun, in which the downwards pumping of the convection is much stronger in 
$z>0$ than in $z<0$. We take $Y_2=0$ and $Y_1<0$ and consider the effects of increasing $|Y_1|$. Flux transport in the 
tachocline region is then controlled only by diffusion. 

\subsubsection{Uniform diffusion}

\label{section:parkerbddg20eqn}

\begin{figure}
\begin{center}
\resizebox{0.95\textwidth}{!}{\includegraphics{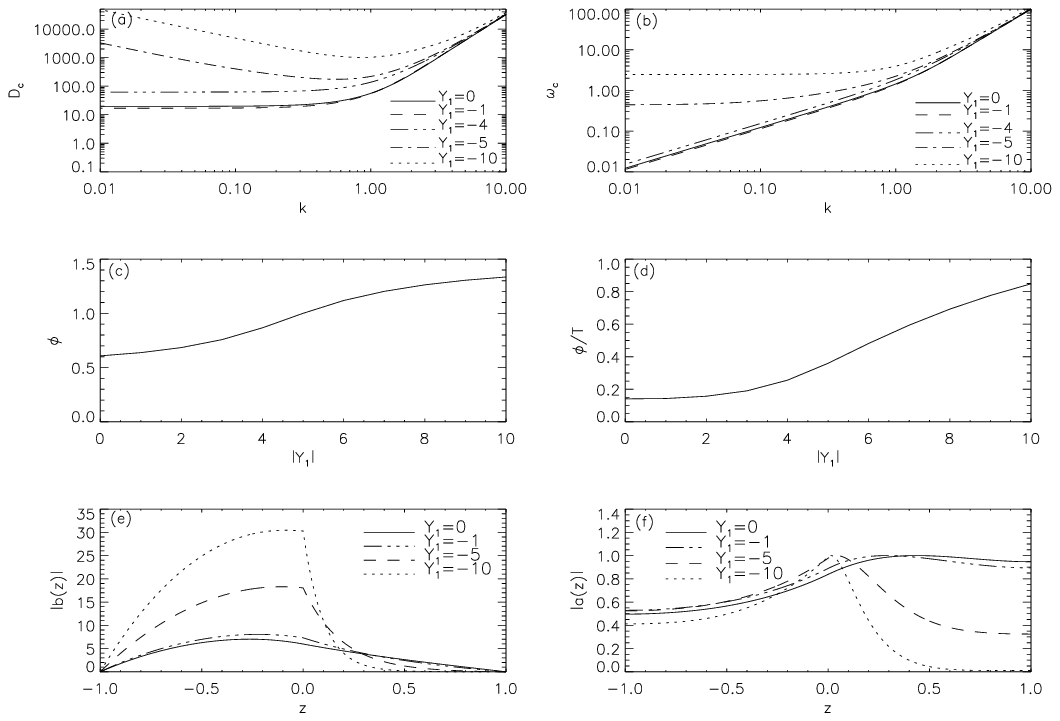}}
\end{center}
\caption{As for Figure~\ref{fig:m3_gud_mu1} with $Y_1<0$, $Y_2=0$, $\eta_1=\eta_2=1$. The toroidal field has been
divided by factors of 2 and 4 for $Y=-5$ and $Y=-10$ respectively.}
\label{fig:m3_g20_mu1}
\end{figure}

We first consider the case of uniform diffusion, $\eta_1=\eta_2=\eta$. Figures~\ref{fig:m3_g20_mu1}(a,b) show that 
there is a magnitude of the transport ($Y_{1_s}(\eta)$, say) for which there is a qualitative change in the 
behaviour of the solution at small wavenumbers. This behaviour is similar to the case of a unidirectional $\gamma$-effect 
(see \S~\ref{section:Parkerbddeqneqg}) although here the results are dependent on the direction of flux transport. For 
$\eta=1$ and $Y_1<0$ we have $Y_{1_s}(1) \approx -4.9$. For $|Y_1|<|Y_{1_s}(\eta)|$ the most readily destabilised 
mode satisfies $D_c \approx D_0(Y_1, \eta)$ and $\omega_c \approx \omega_1(Y_1,\eta)k$ for $k \ll 1$, while for 
$|Y_1| >|Y_{1_s}(\eta)|$ and $k \ll 1$ we have $D_{c} \approx D_{-1}(Y_{1},\eta)k^{-1}$ and 
$\omega_{c} \approx \omega_{0}(Y_{1},\eta)$. Most notably however, and in contrast to all previous calculations, 
it is found that the minimum of $D_c$ over all wavenumbers as a function of the transport speed $Y_1$ possesses a 
stationary value. Thus there exists a preferred magnitude of transport for which dynamo action is most 
efficient. For $\eta=1$ the transport that leads to the most efficient dynamo is $Y_{1}^* \approx -1.2$, where 
$D_c \approx 16.8$. We shall return to this point in \S\ref{section:efficient} [see also Figure~\ref{fig:m3_minDc}(c)].  

The effect of the transport mechanism on the phase relation between the radial and azimuthal fields is illustrated in 
Figures~\ref{fig:m3_g20_mu1}(c,d). The monotonically increasing behaviour with $|Y_1|$ is similar to that for the 
same case in Model~I (cf.~Figures~\ref{fig:m2_g20_mu1}(c,d)).

Figures~\ref{fig:m3_g20_mu1}(e,f) illustrate the structure of the poloidal and toroidal fields as the transition from transport 
enhancing to hindering the dynamo occurs. Although it is difficult to infer from the eigenfunctions alone why a small magnitude 
of transport helps the dynamo (there appears to be little difference in the spatial profile of the solutions in the cases 
$Y_1=0,-1$), it is interesting to note that in the case when the transport exceeds the preferred magnitude the majority of both 
the toroidal and poloidal fields resides in $z<0$. This behaviour is suggestive that $Y_1^*$ corresponds to the strength below 
which the main effect is to enhance the downwards transport of flux into the tachocline,  more readily feeding the 
$\omega$-effect and leading to more efficient dynamo action, whereas above $Y_1^*$ the diffusive transport of toroidal field 
into the convection zone is hindered significantly, thus inhibiting the operation of the $\alpha$-effect and making the dynamo 
cycle more difficult to complete. Interestingly, the behaviour of the fields in $z>0$ bears some similarity to the case of a 
uniform downwards transport throughout the domain [cf.~Figures~\ref{fig:m3_eqg_mu1}(c,d)].  Thus the reason that the 
dynamo is more difficult to excite in the case of uniform transport (the minima of $D_c$ over all $k$ in 
Figure~\ref{fig:m3_eqg_mu1}(a) are greater than those in Figure~\ref{fig:m3_g20_mu1}(a)) is likely to be due to the behaviour 
in $z<0$: when $Y_2$ is finite the toroidal field is translated towards the lower boundary where it is annihilated.

\subsubsection{Reduced diffusion in $z<0$}

\begin{figure}
\begin{center}
\resizebox{0.95\textwidth}{!}{\includegraphics{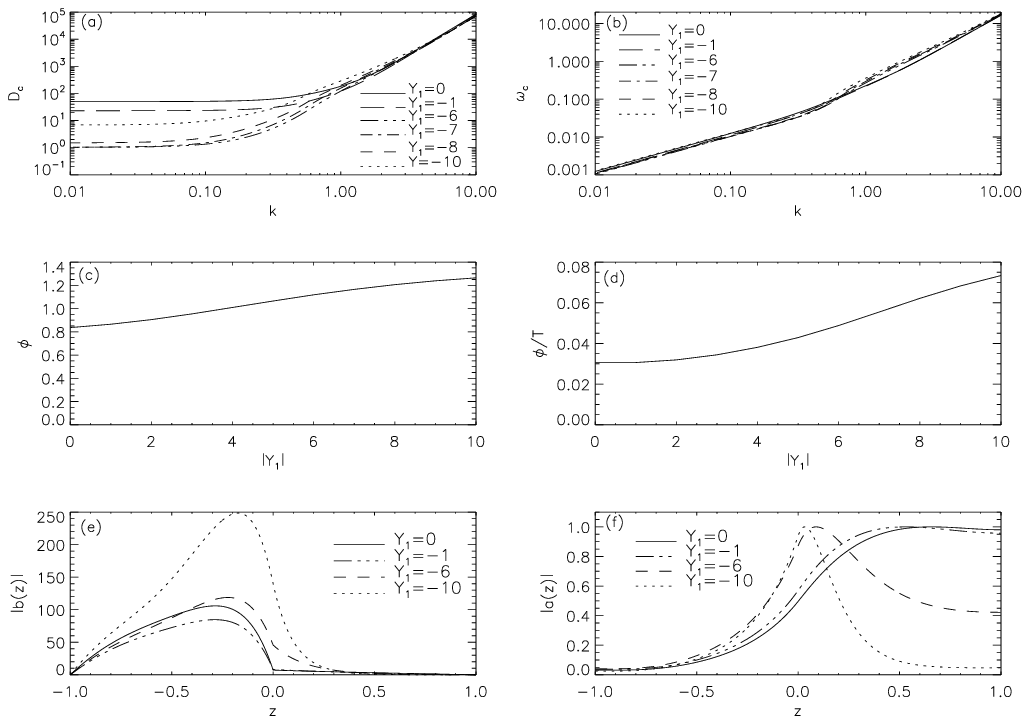}}
\end{center}
\caption{As for Figure~\ref{fig:m3_gud_mu1} with $Y_1<0$, $Y_2=0$, $\eta_1=1$, $\eta_2=0.01$.}
\label{fig:m3solar}
\end{figure}

We now consider the case in which the the diffusivity in $z<0$ is much smaller than that in $z>0$. We take $\eta_1=1$, 
$\eta_2=0.01$ and compare the results with the uniform diffusion case above (\S\ref{section:parkerbddg20eqn}). 
The profiles of the critical dynamo number and corresponding frequency are shown in Figures~\ref{fig:m3solar}(a,b). 

At large values of the wavenumber the profiles collapse onto those for $Y_1=Y_2=0$. Indeed, for $k \gg 1$ and 
$\mu^2 \rightarrow 0$ the critical dynamo number and frequency are given by the relations~(\ref{eq:m2_g0smu_dcwc}) 
(Parker 1993). The numerical results show that for $k \ll 1$, $D_{c} \approx D_{0}(Y_{1}, \eta_1,\eta_2)$ and 
$\omega_{c} \approx \omega_{1}(Y_{1}, \eta_1,\eta_2)k$,
where the coefficients $D_0$ and $\omega_1$ are non-monotonic in $Y_1$. Thus again a preferred magnitude of 
transport, $Y_1^*$, exists for which the dynamo is most easily excited. For $\eta_2=0.01$ and $Y_1<0$, $Y_{1}^* \approx 
-6.5$ [see also Figure~\ref{fig:m3_minDc}(d)]. It is noted that when transport is absent the dynamo is easier to excite when 
$\eta_2=1$ than when $\eta_2=0.01$; however a 
comparison of the preferred strengths in the two cases (i.e.\ $D_{c}(Y_{1}^* \approx -6.5,\eta_2=0.01)=0.99$ 
compared with $D_{c}(Y_{1}^* \approx -1.2,\eta_2=1)=16.8$) shows that the stronger $\gamma$-effect and smaller 
diffusivity ratio results in a more efficient dynamo.

The effect of transport on the phase relation between the radial and azimuthal fields, 
and also the ratio of the phase lag to the period of the wave, is shown in Figures~\ref{fig:m3solar}(c,d). The behaviour is 
qualitatively similar to that for the case of uniform diffusion with $Y_2=0$ (cf.~Figures~\ref{fig:m3_g20_mu1}(c,d)).

Figures~\ref{fig:m3solar}(e,f) illustrate the effect on the eigenfunctions of increasing the transport in $z>0$. The structure of 
the toroidal field is affected only weakly, whereas the poloidal field becomes more localised around the interface, the effect 
being particularly pronounced in $z>0$.

\subsection{The most efficient dynamo}
\label{section:efficient}

\begin{figure}
\begin{center}
\resizebox{0.95\textwidth}{!}{  \includegraphics{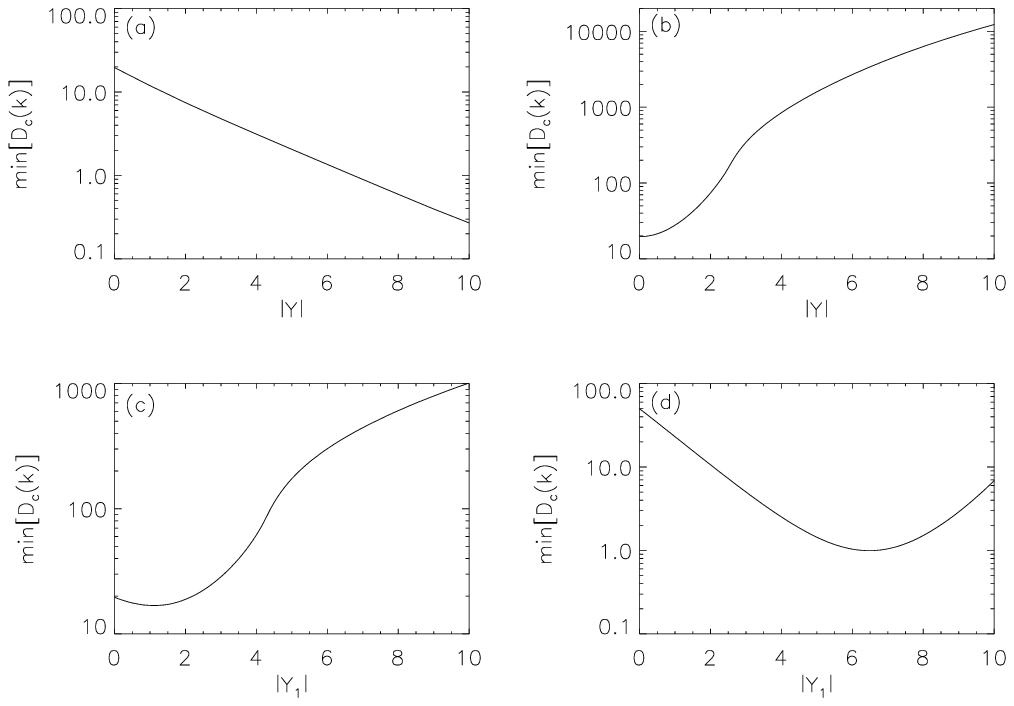}}
\end{center}
\caption{Model~II: min$_k[D_{c}(k)]$ versus magnitude of transport. (a) Bidirectional transport towards the interface with 
uniform 
diffusion ($Y_1=-Y_2=Y$, $Y<0$, $\eta_1=\eta_2=1$), (b) Unidirectional transport and uniform diffusion 
($Y_1=Y_2=Y$, $\eta_1=\eta_2=1$), (c) Dominant transport in $z>0$ with uniform diffusion ($Y_1<0$, $Y_2=0$, 
$\eta_1=\eta_2=1$), (d) Dominant transport in $z>0$ with reduced diffusion in $z<0$ ($Y_1<0$, $Y_2=0$, 
$\eta_1=1$, $ \eta_2=0.01$). Cases (c) and (d) have a finite most efficient strength of the $\gamma$-effect, 
$Y_{1}^* \approx -1.2,-6.5$, respectively.} 
\label{fig:m3_minDc}
\end{figure}

Finally, to summarise the findings for this model and to compare the efficiency of the four cases considered, 
Figures~\ref{fig:m3_minDc}(a-d) present the minimum value of the critical dynamo number over all wavenumbers 
as a function of the $\gamma$-effect for each case. Figure~\ref{fig:m3_minDc}(a) illustrates the most efficient 
dynamo model in which the $\gamma$-effect is directed towards the interface from both above and below. Increasing 
the strength of the $\gamma$-effect in this case monotonically decreases the threshold for the dynamo instability. 
Figure~\ref{fig:m3_minDc}(b) illustrates the least efficient dynamo model. In this case a uniform transport mechanism 
operating throughout the domain hinders dynamo action, with the minimum critical dynamo number increasing monotonically 
with increasing strength of the $\gamma$-effect. Figures~\ref{fig:m3_minDc}(c,d) illustrate the cases when the non-diffusive 
transport is restricted to operate only in $z>0$, with uniform diffusion in the former and a reduced diffusion in $z<0$ in the 
latter. In both of these cases there exists a preferred magnitude of the $\gamma$-effect for which dynamo action is most 
efficient.


\section{Conclusion}
\label{sec:4}

We have considered the role of anisotropic transport on the efficiency of an interface dynamo in 
the linear regime, and its effect in modifying the temporal structure of the solution, through 
the frequency of the dynamo wave and the phase relation between the radial and azimuthal fields. One of the main 
findings of the investigation is that the conclusions to be drawn depend crucially upon the geometry of the model. 

The first model considered was of infinite vertical extent. In the absence of transport, the preferred wavenumber for the onset 
of the dynamo instability is $k=0$, where the critical parameter $(\alpha G/\eta_1^2)_c=0$. Although, clearly, this cannot be 
decreased by incorporating the $\gamma$-effect, for all forms of transport investigated it is found that $\textrm{min}_k 
[(\alpha G/\eta_1^2)_c]$ increases as the magnitude of the transport increases. Thus dynamo action is less readily achieved. 
Also, the period of the dynamo wave decreases 
as the magnitude of the $\gamma$-effect increases. Thus although the dynamo is more difficult to excite, it more quickly 
reverses the polarity of the field. Perhaps most noteworthy is that for this model the two cases of the $\gamma$-effect directed 
towards the interface from both above and below and of the transport directed away from the interface in both regions have 
identical critical dynamo numbers and corresponding frequencies  (\S\ref{sec:m2_bidirectional}).

The effect of the transport mechanism in the more realistic Model~II is more complex. The model geometry is 
the same as that of 
Model~I except that the domain of the problem is bounded in the vertical direction. Here it is found that a uniform 
transport velocity throughout the convection zone and tachocline delays the onset of the dynamo instability and 
increases the frequency of the dynamo waves. However, transport directed towards the interface in 
both $z>0$ and $z<0$ decreases both the critical dynamo number and the frequency. Thus transport aids the onset of the 
instability in this case, although it then takes longer to reverse the field polarity. Restricting the transport mechanism to the 
convection zone reveals a preferred strength for dynamo action.

It is of interest to relate our studies to previous investigations of the effects of transport on dynamo models. { Brandenburg et 
al.~(1993) and Gabov et al.~(1996) considered models of galactic dynamos and investigated the role of turbulent diamagnetism 
on the dynamo instability via the additional term $\nabla \times (-\nabla \beta/2 \times \mathbf{B})$ in the induction equation, 
where $\beta$ is the turbulent diffusivity. Brandenburg et al.~(1993) illustrated that a galactic wind can enhance the instability by 
transporting field into regions of stronger generation, while turbulent diamagnetism can act to oppose this effect and hence 
hinder dynamo action. Similarly, Gabov et al.~(1996) illustrated that the critical dynamo number is larger when the effects of 
turbulent diamagnetism are present. This was attributed, in part, to an increase in the effective diffusion coefficient. It was then 
shown that the effective dynamo number $D_c/\tilde\beta^2$ (where $\tilde\beta$ is the mean value of the turbulent diffusion 
coefficient) decreases as $\tilde\beta$ is increased. It was further illustrated that for highly supercritical dynamo numbers the 
linear growth rate is larger when turbulent diamagnetism is present than in its absence. In our study we have concentrated on the 
effects of transport on marginally stable solutions. Nightingale (1983) incorporated the $\gamma$-effect into a spherical 
$\alpha^2$-dynamo model and found that the dynamo growth rate is smaller when $\gamma$ is non-zero than when there is no 
transport. R\"adler (1986) considered spherical $\alpha\omega$- and $\alpha^2$-dynamos and showed that although radial 
transport mainly hinders dynamo action, certain strengths of $\gamma$ can promote the instability.}

Finally, we remark that interface dynamo models relying purely upon diffusive communication are sometimes rejected owing to 
their inability to reproduce either the observed direction of propagation of the dynamo belts (exemplified in the butterfly 
diagram) or the observed phase relation between the radial and azimuthal fields. In this regard, it is of interest to note that the 
study by Brandenburg et al. (1992) showed that a sufficiently strong downward pumping can lead to an equatorward migration 
of the magnetic field even when the conventional rule for the sign of $\alpha \partial \Omega/\partial r$ yields a poleward 
migration. Also, here we have shown that the transport mechanism changes the ratio of the phase difference to the period of the 
dynamo wave. It is therefore entirely possible that tuning the $\gamma$ parameter could enable interface dynamo models to 
reproduce the observed behaviour.

\section*{Acknowledgements}
\label{sec:5}
J.M. was funded by a Ph.D studentship from the EPSRC, UK, and by the NSF sponsored Center for Magnetic 
Self-Organization at the University of Chicago. The paper was completed at the Kavli Institute for Theoretical Physics, 
supported by Grant No.\ NSF PHY05-51164.

\section*{Appendix A -- Dispersion relation for Model~I}
\label{sec:appendix_a}

In $z>0$ we seek  solutions  of the form
\begin{equation}
\label{eq:B1}
B_{1}=C\exp{[t(\sigma+\textrm{i}\omega)]}\exp{[-z(S+\textrm{i}Q)]}\exp{(\textrm{i}kx),}
\end{equation}
\begin{equation}
\label{eq:A1}
A_{1}=(F+Ez)\exp{[t(\sigma+\textrm{i}\omega)]}\exp{[-z(S+\textrm{i}Q)]}\exp{(\textrm{i}kx),}
\end{equation}
where $C$, $E$, $F$ are constants and $\sigma$, $\omega$, $S$, $Q$, $k$ are
real, with \(S>0\) so that \(\boldsymbol{B} \rightarrow 0\) as
\(z\rightarrow + \infty\). Substitution into equations (\ref{eq:dynamo_z>0})
gives
\begin{equation}
\label{eq:E} E[2\eta_{1}(S+\textrm{i}Q)+\gamma_{1}]=\alpha C, \qquad \sigma+\textrm{i} \omega = -k^{2}\eta_{1} +
\eta_{1}(S+\textrm{i}Q)^{2} + \gamma_{1} (S+\textrm{i}Q),
\end{equation}
with $F$ arbitrary.

In \(z<0\) we look for solutions of the form
\begin{equation}
\label{eq:A2}
A_{2}=J \exp{[t(\sigma+\textrm{i}\omega)]}\exp{[z(s+\textrm{i}q)]}\exp{(\textrm{i}kx),}
\end{equation}
\begin{equation}
\label{eq:B2}
B_{2}=(L+Mz)\exp{[t(\sigma+\textrm{i}\omega)]}\exp{[z(s+\textrm{i}q)]}\exp{(\textrm{i}kx),}
\end{equation}
where $J$, $L$, $M$ are constants and $\sigma$, $\omega$, $s$, $q$, $k$ are
real with \(s>0\) so that \(\boldsymbol{B} \rightarrow 0\) as
\(z\rightarrow - \infty\). Substitution into equations (\ref{eq:dynamo_z<0})
then gives
\begin{equation}
\label{eq:sigmaomega<} M[2\eta_{2}(s+\textrm{i}q)-\gamma_{2}]=-\textrm{i}kGJ, \qquad \sigma+\textrm{i} \omega = 
-k^{2}\eta_{2} +
\eta_{2}(s+\textrm{i}q)^{2} - \gamma_{2} (s+\textrm{i}q),
\end{equation}
with $L$ arbitrary. Applying continuity of $A$ and $B$ across $z=0$, together with the jump conditions
(\ref{eq:jump}), yields
\begin{equation}
\label{eq:FJ} F=J, \qquad C=L, \qquad
\gamma_{2}L-\gamma_{1}C-\eta_{2}M-\eta_{2}L(s+\textrm{i}q)-\eta_{1}(S+\textrm{i}Q)C=0, \qquad 
E=F(S+\textrm{i}Q)+J(s+\textrm{i}q).
\end{equation}
Equations (\ref{eq:E}a), (\ref{eq:sigmaomega<}a) and (\ref{eq:FJ}) may be combined to yield two equations
for $C$ and $F$. A non-trivial solution requires that the determinant of the coefficients of $C$ and $F$ vanishes, from which 
we obtain the relation
\begin{eqnarray}
\label{eq:disp}
\left( S+\textrm{i}Q+s+\textrm{i}q \right) 
\left( (S+\textrm{i}Q)+\mu^2(s+\textrm{i}q)
+ \frac{(\gamma_{1}-\gamma_{2})}{\eta_{1}}\right)
\left( 2(S+\textrm{i}Q)
+\frac{\gamma_{1}}{\eta_{1}} \right)
\left( 2(s+\textrm{i}q)-\frac{\gamma_{2}}{\eta_{2}} \right) = 
\textrm{i}k^4D,
\end{eqnarray}
where $D=\alpha G  / \eta_{1}^2 k^3$ and $\mu^2=\eta_{2}/\eta_{1}$. Recall that we also have equations (\ref{eq:E}b) and
(\ref{eq:sigmaomega<}b) that relate $s$, $q$, $S$ and $Q$ to the complex growth rate.

\section*{Appendix B -- Dispersion relation for Model~II}

We consider first the convection zone ($z>0$). The dynamo equations (\ref{eq:nondima}), together with the proposed form of 
solutions (\ref{eq:solns}), yield
\begin{equation}
\label{eq:bz>0}
\frac{d^2 b_{1}}{dz^2}-\frac{Y_{1}}{\eta_{1}}\frac{db_{1}}{dz} -\frac{q_{1}^2}{\eta_{1}}b_{1} =0, \qquad
\frac{d^2a_{1}}{dz^2}-\frac{Y_{1}}{\eta_{1}}
\frac{da_{1}}{dz}-\frac{q_{1}^2 a_{1}}{\eta_{1}}=- \frac{\alpha
b_{1}}{\eta_{1}},
\end{equation}
where $q_{1}^2=p+ \eta_{1} k^2$ and $Y_1 = R_{\gamma} \gamma_1$. Solving equation (\ref{eq:bz>0}a) for \(b_{1}\)
and applying boundary condition (\ref{eq:boundary1ab}a) yields
\begin{equation}
\label{eq:b1} 
b_1(z)=2A \textrm{sinh}[ \overline m (z-1)] 
\exp{\left(\frac{Y_{1}(z-1)}{2 \eta_{1}}\right)},
\end{equation}
where $A$ is a constant and $\overline m = \sqrt{Y_{1}^2+4 \eta_{1} q_{1}^2}/ 2 \eta_{1}$. Substituting $b_{1}$ into
equation (\ref{eq:bz>0}b) and solving the
homogeneous equation we obtain
\begin{equation}
a_{1_{c}}(z) = \exp{\left( \frac{Y_{1} (z-1)}{2 \eta_{1}} \right)} 
\left( C \textrm{cosh} [ \overline m (z-1)] + 
E \textrm{sinh}[ \overline m (z-1)] \right). 
\end{equation}
Looking for a  particular solution of the inhomogeneous part of the form
\begin{equation}
a_{1_{p}}(z)= z \exp{ \left( \frac{Y_{1} (z-1)}{2 \eta_{1}} \right) }
\left( I \textrm{cosh}[ \overline m (z-1)]+F \textrm{sinh}[ \overline m (z-1)] \right)
\end{equation}
we find $I=-\alpha A/ \eta_1 \bar m$ and $F=0$. On application of boundary condition (\ref{eq:boundary1ab}b) we finally
obtain the full solution
\begin{eqnarray}
\label{eq:z>0}
a_{1}(z)= \exp{\left( \frac{Y_{1} (z-1)}{2 \eta_{1}} \right) } 
\left( C \textrm{cosh}[ \overline m (z-1)]+ \left[\frac{A \alpha}{ \eta_{1} \overline m^2} 
\left( 1+\frac{Y_{1}}{2 \eta_{1}} \right)
- \frac{Y_{1}C}{2 \eta_{1} \overline m} \right]
\textrm{sinh}[ \overline m (z-1)]- \frac{z A \alpha}{\overline m \eta_{1}} 
\textrm{cosh}[ \overline m (z-1)]\right).
\end{eqnarray}
Following a similar procedure in the tachocline region ($z<0$), the corresponding equations to (\ref{eq:bz>0}) read
\begin{equation}
\label{eq:bz<0}
\frac{d^2 b_{2}}{dz^2}-\frac{Y_{2}}{\eta_{2}}
\frac{db_{2}}{dz} -\frac{q_{2}^2}{\eta_{2}}b_{2} =-\frac{\textrm{i}kDGa_{2}}{\eta_2}, \qquad 
\frac{d^2a_{2}}{dz^2}-\frac{Y_{2}}{\eta_{2}}
\frac{da_{2}}{dz}-\frac{q_{2}^2}{\eta_{2}}a_2=0,
\end{equation}
where \(q_{2}^2=p+ \eta_{2} k^2\), with solutions
\begin{equation}
\label{eq:z<0} 
a_{2} = J \exp{\left( \frac{Y_{2} (z+1)}{2 \eta_{2}} \right) } 
\left( \textrm{cosh}[ \overline n (z+1)] - \frac{Y_{2}}{ 2 \eta_{2} \overline n}
\textrm{sinh}[ \overline n (z+1)] \right) ,
\end{equation}
\begin{equation}
b_{2}(z) = \exp{\left( \frac{Y_{2} (z+1)}{2 \eta_{2}}\right) } 
\left( \frac{\textrm{i}kDGY_{2}J}{4 \eta_{2}^2\overline n^2}(1+z)
\textrm{cosh}[\overline n (z+1)]+\left(L-\frac{\textrm{i}kDGJz}{2 \eta_{2}\overline n}\right) 
\textrm{sinh}[\overline n (z+1)]
\right),
\end{equation}
where $\bar n=\sqrt{Y_2^2+4 \eta_2 q_2^2}/2 \eta_2$. Applying the jump conditions (\ref{eq:jump}) (with $\gamma$
replaced by $Y$) and continuity of $A$ and $B$ across $z=0$ gives four equations for the four unknown constants $A$,
$C$, $J$ and $L$. A non-trivial solution requires that the determinant of the coefficients vanishes, from which we obtain the 
dispersion relation
\begin{eqnarray}
&&-2(2\eta_2\bar n\cosh[\bar n]-Y_2\sinh[\bar n])\Big[\textrm{i}\alpha DGY_1Y_2k\bar m\cosh^2[\bar m]\cosh[\bar 
n]+\nonumber \\
&& \cosh[\bar m]\sinh[\bar m]\big(-\textrm{i}\alpha DG Y_1 Y_2k\cosh[\bar n]+4\eta_1^2\bar m^2 \bar n(-Y_2^2+4\eta_2^2 
\bar
n^2)\sinh[\bar n]\big)-\nonumber \\ 
&& \bar m\sinh^2[\bar m]\Big(\big[\textrm{i}\alpha DG(2\eta_1+Y_1)Y_2 k+4\eta_2^2(Y_1^2-4\eta_1^2 \bar m^2)\bar 
n^2\big]\cosh[\bar n]\nonumber \\
&& +2\bar n\big(\eta_1 Y_1 Y_2^2-\eta_2 Y_2(Y_1^2-4\eta_1^2 \bar m^2)-4\eta_1\eta_2^2 Y_1 \bar n^2\big)\sinh[\bar 
n]\Big)\Big]+\nonumber \\
&& \sinh[\bar n]\Big\{\textrm{i}\alpha DG(2\eta_2-Y_2)k\big(-2(\eta_1+ Y_1)\bar m+2\eta_1 \bar m\cosh[2\bar 
m]+Y_1\sinh[2 \bar 
m]\big)(-Y_2\cosh[\bar n]+2\eta_2 \bar n\sinh[\bar n])\nonumber \\ 
&& -4\bar m \bar n(2\eta_1 \bar m\cosh[\bar m]+Y_1\sinh[\bar m])\Big[-2\eta_2^2(Y_1^2-4\eta_1^2 \bar m^2)\bar 
n\cosh[\bar n]\sinh[\bar m]+\nonumber \\
&& \Big(-2\eta_1^2 \bar m(Y_2^2-4\eta_2^2\bar n^2)\cosh[\bar m]+\big(-\eta_1 Y_1 Y_2^2+\eta_2 Y_2(Y_1^2-4\eta_1^2 
\bar m^2)+4\eta_1\eta_2^2 Y_1 \bar n^2\big)\sinh[\bar m]\Big)\sinh[\bar n]\Big]\Big\}=0
\label{eq:m3_disp}
\end{eqnarray}

\end{document}